\documentclass[12pt]{article}
\usepackage{graphicx}
\usepackage{overpic}
\usepackage[mathlines]{lineno}

\def\lbnl{Physics Division, Lawrence Berkeley National Laboratory, Berkeley, CA 94720, USA}

\def\Title#1{\begin{center} {\Large #1 } \end{center}}
\def\Author#1{\begin{center}{ \sc #1} \end{center}}
\def\Address#1{\begin{center}{ \it #1} \end{center}}

\newenvironment{Abstract}{\begin{quotation}  }{\end{quotation}}
\newenvironment{Presented}{\begin{quotation} \begin{center}
             PRESENTED AT\end{center}\bigskip
      \begin{center}\begin{large}}{\end{large}\end{center} \end{quotation}}

\def\beq{\begin{equation}}
\def\eeq#1{\label{#1}\end{equation}}
\def\eeqn{\end{equation}}

\def\beqa{\begin{eqnarray}}
\def\eeqa#1{\label{#1}\end{eqnarray}}
\def\eeqan{\end{eqnarray}}

\let\bar=\overbar

\def\Dslash{\not{\hbox{\kern-4pt $D$}}}
\def\dslash{\not{\hbox{\kern-2pt $\del$}}}

\def\msb{{\bar{\ssstyle M \kern -1pt S}}}

\begin{document}

\begin{titlepage}

\vfill
\Title{Expected performance of the upgrade ATLAS experiment\\ for HL-LHC}
\vfill
\Author{ Peilian Liu on behalf of the ATLAS Collaboration}
\Address{\lbnl}
\vfill
\begin{Abstract}
The Large Hadron Collider (LHC) has been successfully delivering proton-proton collision data at the unprecedented center of mass energy of 13\,TeV. An upgrade is planned to increase the instantaneous luminosity delivered by the LHC in what is called the HL-LHC, aiming to deliver a total of up 3000\,fb$^{-1}$ to 4000\,fb$^{-1}$ of data per experiment. To cope with the expected data-taking conditions ATLAS is planning upgrades of the detector.
Six Technical Design Reports (TDR) were produced by the ATLAS Collaboration. In these TDRs the physics motivation and benefits of such upgrades are discussed together with details on the upgrade project itself. In this contribution we review the expected performance of the upgraded ATLAS detector and the expected reach for physics measurements as well as the discovery potential for new physics that is expected by the end of the HL-LHC data-taking. The performance of object reconstruction under the expected pile-up conditions will be shown, including a fully re-optimized $b$-tagging algorithm. Important benchmark physics projections including di-Higgs boson production sensitivity will be discussed.
\end{Abstract}
\vfill
\begin{Presented}
Thirteenth Conference on the Intersections of Particle and Nuclear Physics (CIPANP2018)\\
Palm Spring, CA, USA,  May 29 -- June 03, 2018
\end{Presented}
\vfill

\insert\footins{\noindent\footnotesize\textcopyright {2018 CERN for the benefit of the ATLAS Collaboration. \\
Reproduction of this article or parts of it is allowed as specified in the CC-BY-4.0 license.}}

\end{titlepage}

\def\thefootnote{\fnsymbol{footnote}}
\setcounter{footnote}{0}

\section{Introduction}
The high-luminosity phase of the Large Hadron Collider (HL-LHC) is foreseen to start in 2026 and will deliver an integrated luminosity of up to 4000\,fb$^{-1}$ in about 10 years.
The HL-LHC will operate at an instantaneous luminosity up to $\mathcal{L}=7.5\times10^{34}$\,cm$^{-2}$s$^{-1}$ which corresponds to an average $<\mu>=200$ inelastic proton-proton collisions per bunch-crossing (pile-up).

The HL-LHC will offer us the opportunities to precisely measure the Higgs coupling using as many Higgs production and decay channels as possible. These precision measurements will provide constraints on potential non-Standard Model physics. With the 4000\,fb$^{-1}$ of data, it allows the exploration of Higgs potential by studying Higgs-boson pair production. The high luminosity will also extend the reach of many new physics searches.

The HL-LHC will present an extremely challenging environment to the ATLAS experiment, well beyond that for which it was designed. An upgrade of the ATLAS detector~\cite{ATLAS-detector} will be needed before the start of this new phase to cope with the high-radiation environment and the large increase in pile-up.

\section{Upgrades of ATLAS detector for HL-LHC}
To reach the physics goals at HL-LHC, maintaining or improving the performance of the ATLAS detector is essential. The ATLAS detector will be upgraded to cope with the increased occupancies and data rates. The inner tracking detector will be completely replaced with a new all-silicon tracker. The Transition Radiation Tracker (TRT) will not be used any more. For the Muon detector, new inner barrel chambers will be installed for trigger.  The Calorimeters themselves are expected to operate reliably during the HL-LHC period, but the readout system of all calorimeters will be upgraded. Actually all readout has to be upgraded because the trigger rate will be increased by an order of magnitude. In addition, the high-granularity timing detector has been proposed and approved. A possible high-$\eta$ muon tagger is under consideration.

This proceeding focuses on few aspects. More details are in Technical Design Reports (TDRs). TDRs for Pixel~\cite{pixel-TDR}, Strip~\cite{strip-TDR}, Muon~\cite{Muon-TDR}, Liquid Argon Calorimeter~\cite{LAr-TDR}, Tile Calorimeter~\cite{Tile-TDR} and TDAQ~\cite{TDAQ-TDR} upgrade have been approved at ATLAS experiment. The HGTD has produced a technical proposal~\cite{HGTD-TP} and the TDR is planned for early 2019.

\subsection{New Inner Tracker}
\begin{figure}[h]
\centering
\includegraphics[width=0.7\textwidth]{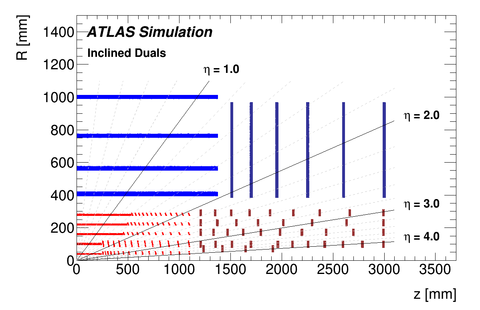}
\caption{A schematic layout of the ITk Inclined Duals layout for the HL-LHC phase of ATLAS~\cite{pixel-TDR}. Here only one quadrant and only active detector elements are shown. The horizontal axis is the axis along the beam line with zero being the interaction point. The vertical axis is the radius measured from the interaction region. The outer radius is set by the inner radius of the barrel cryostat that houses the solenoid and the electromagnetic calorimeter.}
\label{fig:ITk-layout}
\end{figure}

An all-silicon Inner Tracker (ITk) has been designed to completely replace the current Inner Detector (ID).
As shown in Fig.~\ref{fig:ITk-layout}, the ITk comprises two subsystems: a Strip Detector surrounding a Pixel Detector. The Strip Detector has four barrel layers and six end-cap petal-design disks on each side, both having double modules each with a small stereo angle to add $z$ $(R)$ resolution in the barrel (end-caps). The Strip Detector, covering $|\eta|<2.7$, is complemented by a five layer Pixel Detector extending the coverage to $|\eta|<4$. Radiation-hard sensors and readout chips that can withstand the particle fluence have been designed. High-granularity pixels are necessary to reduce the occupancy and handle the high pile-up environment. The inclined sensors in barrel reduce the amount of silicon needed by maximizing angular coverage per sensor. End-cap modules are supported in rings instead of traditional disks to simplify mechanics and services routing while optimizing the distribution of hits along high rapidity tracks. The silicon surface is about 200\,m$^2$ with $\sim$5\,G pixels and $\sim$80\,M strips.

\subsection{Muon Spectrometer}
\begin{figure}[h]
\centering
\includegraphics[width=0.7\textwidth]{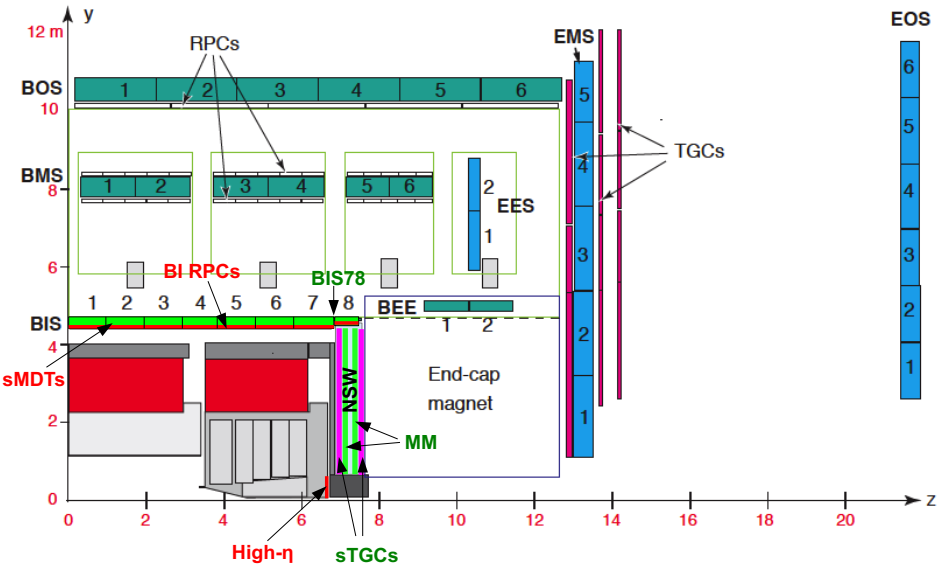}
\caption{$R$-$z$ view of the Phase-II ATLAS muon spectrometer layout~\cite{Muon-TDR}. The drawings show the new detectors to be added in the Phase-II upgrade (red text), those to be installed during Phase-I upgrade (green text), and those that will remain unchanged from the Run 1 layout (black text). }
\label{fig:muon-upgrade}
\end{figure}

The upgrades to the Muon Spectrometer will be done in two phases. All the upgrades for phase-I and II are shown in Fig.~\ref{fig:muon-upgrade}. In the Phase-I upgrade starting in 2019, the present small wheels of the end-cap layers (between toroid and end-cap Calorimeter) will be completely replaced. New Small Wheel (NSW) are required to maintain low $p_{\rm{T}}$ lepton triggers in high-rate environment after Run 2. It could reject up to 90\% of fake triggers at high rates. In the NSW region, the Micromegas (MM) and small-strip Thin Gap Chambers (sTGC) will take the dual role of trigger and precision chambers. Upgrades will also be done to the inner barrel resistive plate chambers (RPC).

The phase-II upgrade will be mainly about trigger. New inner RPC stations will be installed. And the monitored drift tubes (MDTs) information will be added at the hardware trigger. In addition, a high-$\eta$ tagger is under investigation.

\subsection{High-Granularity Timing Detector}
The pile-up will be one of the main challenges of the HL-LHC, the interaction region will spread over about 50\,mm (RMS) along the beam axis and produce an average of 1.6 collisions/mm for $<\mu>=200$. This is to be compared with the 0.24 collisions/mm in Run 2.
The precise assignment of tracks to hard-scatter vertices is crucial to mitigate the pile-up effects. This depends on the space separation of vertices in the beam direction. It will be very challenging at HL-LHC because of the high density of pile-up. The longitudinal impact parameter ($z_0$) resolution in the forward region is limited by multiple scattering.

\begin{figure}[h]
\centering
\includegraphics[width=0.7\textwidth]{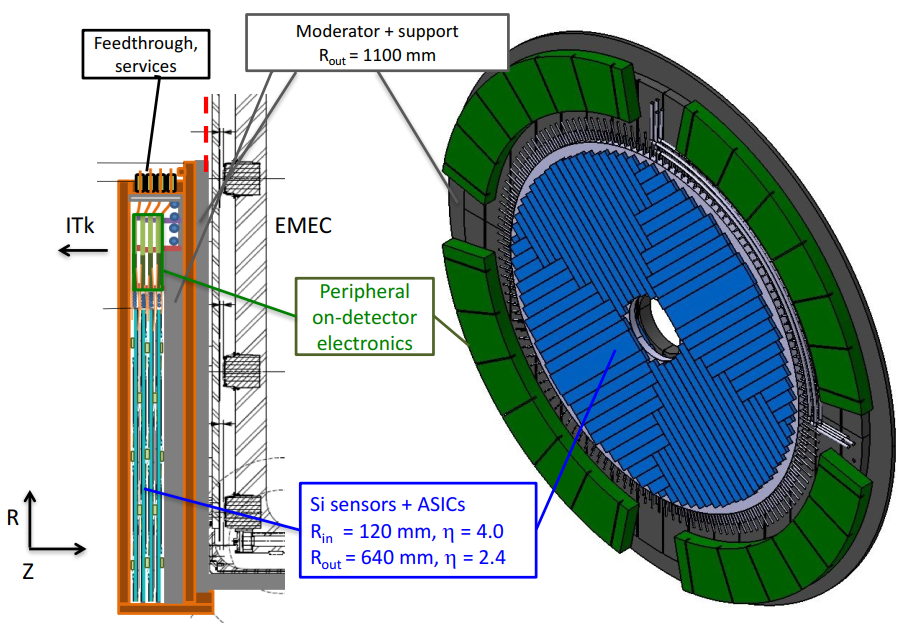}
\caption{The HGTD installed outside of ITk~\cite{HGTD-TP}. }
\label{fig:HGTD}
\end{figure}

A powerful new way to address this challenge is to exploit the time spread of the interactions to distinguish between collisions occurring very close in space but well separated in time. A High-Granularity Timing Detector (HGTD), based on low gain avalanche detector technology, is proposed for the ATLAS Phase-II upgrade. As shown in Fig.~\ref{fig:HGTD}, the HGTD will be located just outside of ITk, covering the pseudorapidity region between 2.4 and 4.0.
In order to ensure a timing resolution of 30\,ps per track over the whole HL-LHC period, four layers of active sensors will be built. At the smallest radius the occupancy is 10\% for a pixel size of $1.3\times1.3$\,mm$^2$.

\subsection{Trigger and DAQ Upgrade}
The high instantaneous luminosity means higher data rates. It poses significant challenges to the Trigger and to the DAQ system to fully exploit the physics potential of the HL-LHC. The baseline architecture is based on a single-level hardware trigger plus event filter. The trigger rate will be 1\,MHz instead of 100\,kHz of Run 2. This will be a big challenge for the detector readout.
Up to 10\,kHz event data are sent into storage. This is ten times of the Run 2 output data rate. To cope with the increased occupancies and high data rates, the readout electronics for all systems will be upgraded.

\section{Expected performance of physics objects}
With the upgrades to the ATLAS detector and the new designed TDAQ system, the physics objects are expected to have a comparable or better performance than Run 2.  The NSW phase-I upgrade and the phase-II upgrades of Muon Spectrometer are not included in the simulation. The new ITK has been included.

\subsection{Tracking Performance}
\begin{figure}[h]
\centering
\includegraphics[width=0.46\textwidth]{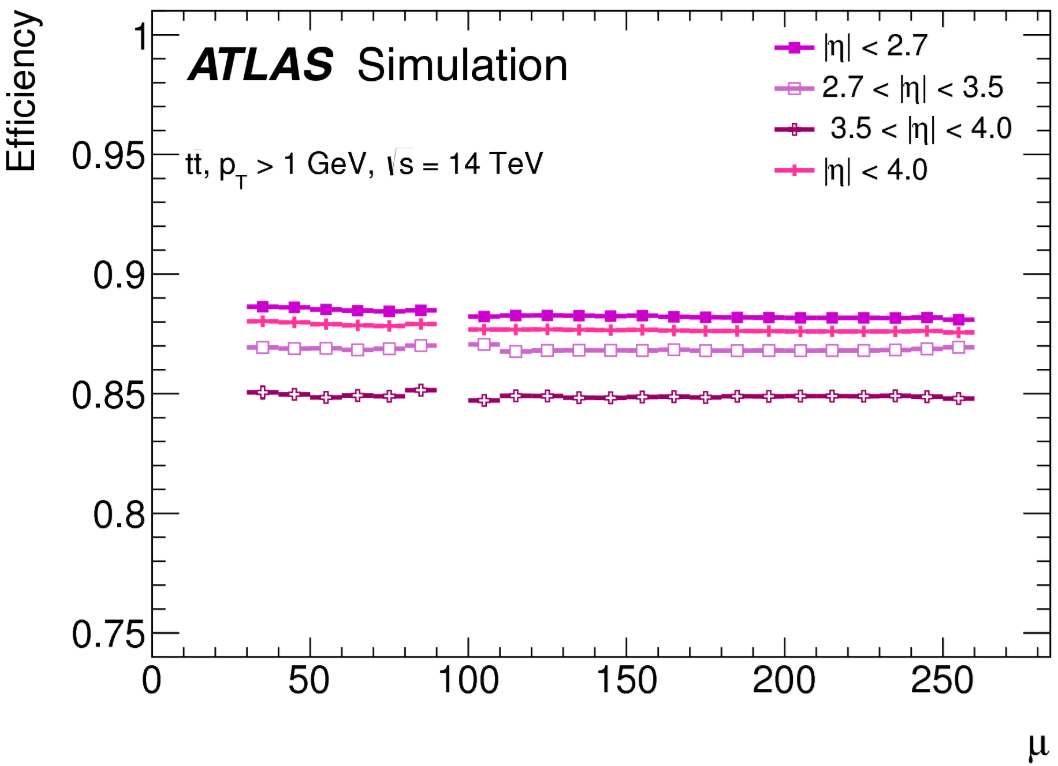}
\hspace{0.5cm}
\includegraphics[width=0.46\textwidth]{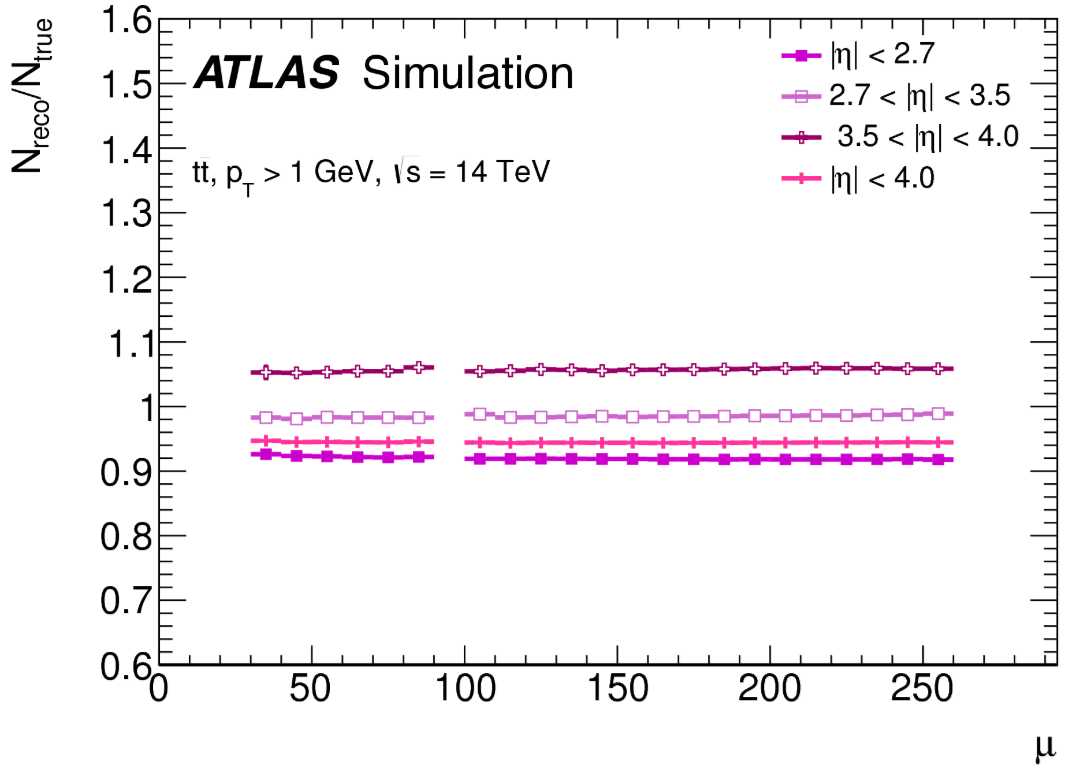}
\caption{Left: Tracking efficiency~\cite{pixel-TDR}. Right: Ratio of reconstructed tracks to generated charged particles~\cite{pixel-TDR}. In both cases, the results are shown as a function of pile-up for $t\bar{t}$ events for pile-up $\mu$ between 40 and 250, for different regions in $\mu$.}
\label{fig:tracking}
\end{figure}

The track reconstruction efficiency in the ITk is found to be stable to better than 1\% for $t\bar{t}$ events with 40 to 250 pile-up, as shown on the left side of Fig.~\ref{fig:tracking} for different regions in $\eta$. For a fixed tracking efficiency, it is important to control the rate of fake or mis-reconstructed tracks. An inclusive measure for the rate of such fake or mis-reconstructed tracks is the ratio of the number of reconstructed tracks to the number of generated true particles.
Because of the limited momentum resolution, the dominant contribution to the additional reconstructed tracks in the forward region is due to mis-measured low-$p_{\rm{T}}$ tracks that have a reconstructed transverse momentum above 1\,GeV.
As can be seen on the right side of Fig.~\ref{fig:tracking}, the ratio does not vary over the full range of pile-up studied between 40 and 250 by more than 1\% for all $\eta$ regions.

\begin{figure}[h]
\centering
\includegraphics[width=0.46\textwidth]{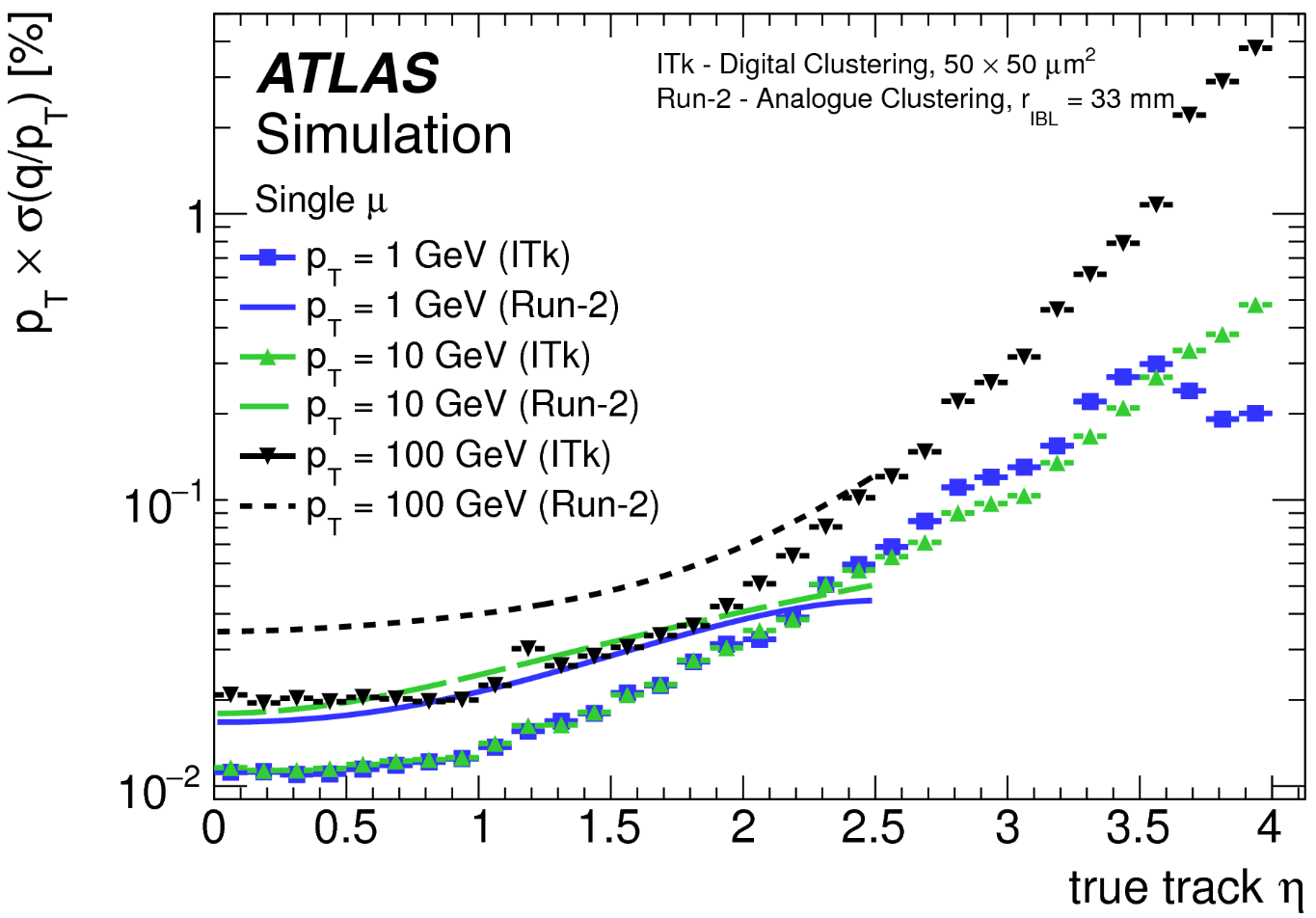}
\hspace{0.5cm}
\includegraphics[width=0.46\textwidth]{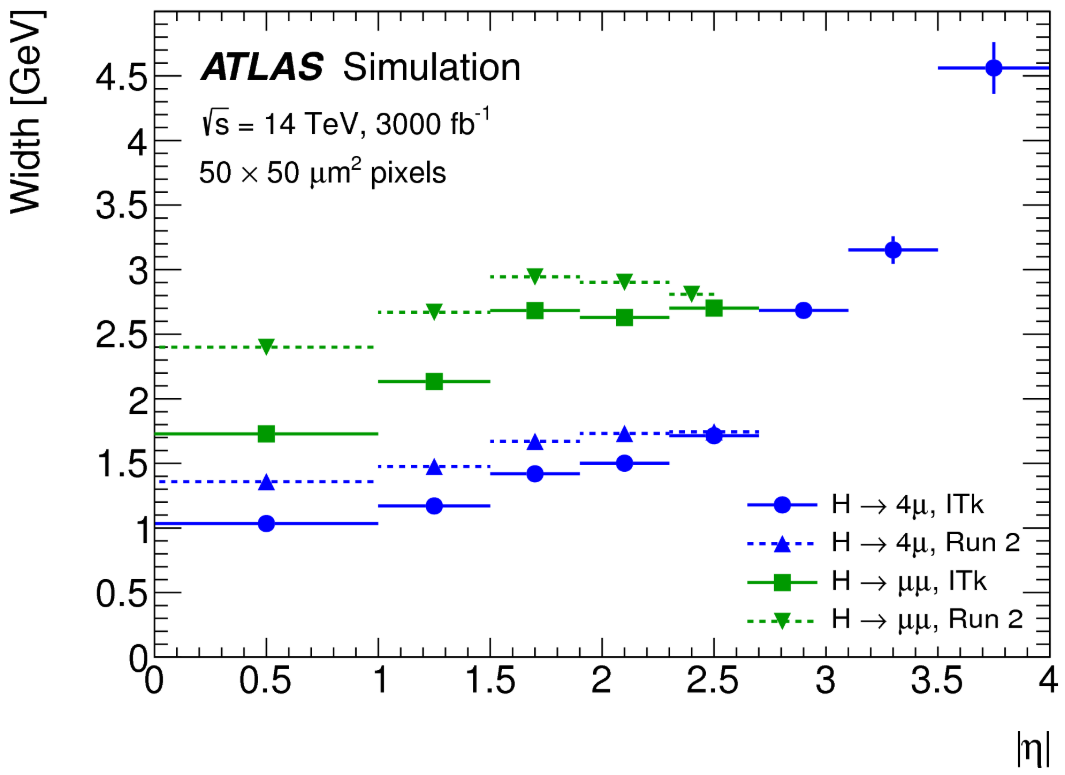}
\caption{(a) Track parameter resolution of $p_{\rm{T}}$ as a function of $\eta$~\cite{pixel-TDR}. (b) The width of the $H\rightarrow ZZ\rightarrow 4\mu$ and $H\rightarrow \mu\mu$ invariant mass signal as a function of the $|\eta|$ of the muon with the largest $|\eta|$ value~\cite{pixel-TDR}.}
\label{fig:trkRes}
\end{figure}

Better track parameter resolution is expected with ITk due to the smaller pixels. The resolution of momentum is shown in the left plot of Fig.~\ref{fig:trkRes} as an illustration. The momentum resolution is nearly a factor 2 better in the ITk than in the Run 2 ID, primarily due to the higher precision of the strip tracker compared to the TRT at high $p_{\rm{T}}$, and due to the reduced material at low $p_{\rm{T}}$. It degrades towards the forward region, mostly due to the fact that the solenoid field in this region is weaker and the transverse path length of the tracks gets shorter. Better momentum resolution leads to better mass resolution as shown in the right plot of Fig.~\ref{fig:trkRes}. The width of the $HH\rightarrow ZZ\rightarrow 4\mu$ and $H\rightarrow \mu\mu$ invariant mass signal have been studied. The resolution is visibly improved with ITk.
The resolution of transverse impact parameter ($d_0$) of ITk is expected to be comparable with Run 2, while the resolution of $z_0$ will be significantly better than Run 2 benefitting from the smaller pixel pitch in the $z$ direction.

The resolutions of these parameters directly determine the performance of the detector in terms of its capability for $b$-tagging and lepton or jet reconstruction.

\subsection{$b$-tagging performance}
The $b$-tagging algorithms used in ATLAS are based on multivariate techniques combing the impact parameters of associated tracks and the properties of reconstructed secondary vertex. The $b$-tagging performance is characterized by the probability to identify jets containing a $b$-hadron decay ($b$-jet efficiency) and by the rejection of jets not containing a $b$- or $c$-hadron as a $b$-jet (light-jet rejection).

\begin{figure}[h]
\centering
\includegraphics[width=0.46\textwidth]{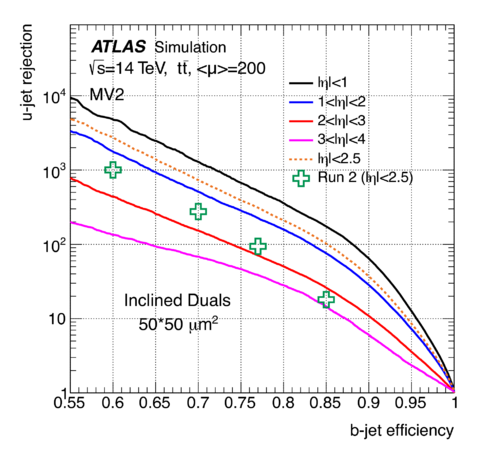}
\caption{Performance of the MV2 $b$-tagging algorithms in $t\bar{t}$ events with 200 pile-up for the ITk layout~\cite{pixel-TDR}. The rejection of light jets for different $\eta$ regions is shown as a function of $b$-jet efficiency. For comparison purposes, the performance for ATLAS during Run 2 with an average of 30 pile-up events is shown as crosses.}
\label{fig:btaggin-ITk-vs-run2}
\end{figure}

The $b$-tagging algorithm called MV2 has been fully re-optimized for the new ITk layout. Fig.~\ref{fig:btaggin-ITk-vs-run2} shows the light-jet rejection versus the $b$-tagging efficiency obtained by $t\bar{t}$ events with an average of 200 pile-up. For comparison purposes, the performance of the Run 2 detector for $|\eta|<2.5$ for events with an average of $\sim$30 pile-up events is also shown. The ITk performs better than the current ID, even at pile-up levels of HL-LHC. Even in the very forward region, the ITk provides significant discrimination power between $b$- and light-jets.

The $b$-tagging is particulary sensitive to the contamination of pile-up tracks. It considers tracks with large impact parameters. So tracks from nearby pile-up are likely to be selected. It is essential to mitigate effects from pile-up.

\subsection{Pile-up Jets Suppression}
\begin{figure}[h]
\centering
\includegraphics[width=0.46\textwidth]{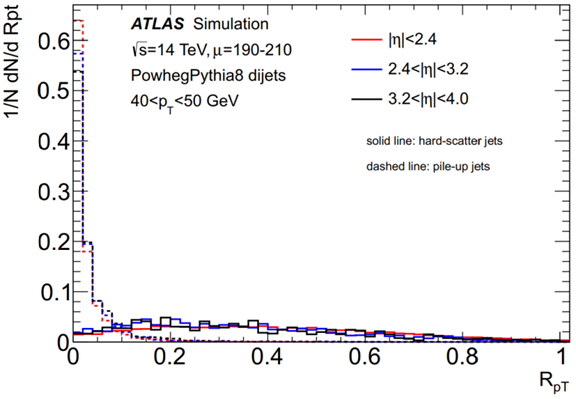}
\hspace{0.5cm}
\includegraphics[width=0.46\textwidth]{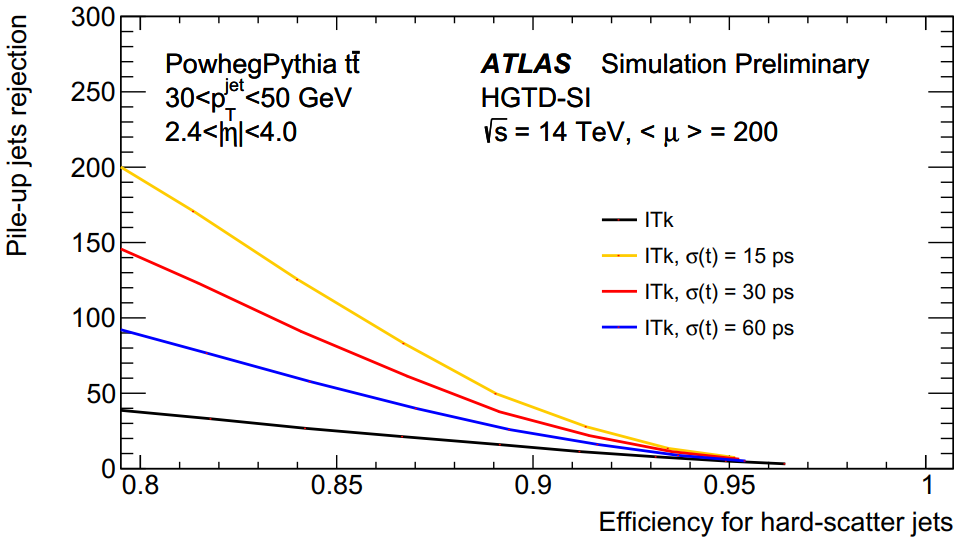}
\caption{(a) $R_{\rm{p_{\rm{T}}}}$ distribution for hard-scatter and pile-up jets with $40<p_{\rm{T}}<50$\,GeV in different $\eta$ regions~\cite{phaseII-scoping}. (b) Pile-up jet rejection as a function of hard-scatter jet efficiency in the $2.4<|\eta|<4.0$ region, for the ITk-only and ITk + HGTD scenarios with different time resolutions~\cite{HGTD-TP}.}
\label{fig:rejection-pileup-vs-eff-HS}
\end{figure}

Pile-up jets are tagged with the discriminant $R_{p_{\rm{T}}}$. It is defined as the scalar $p_{\rm{T}}$ sum of the tracks that are associated with a jet and originate from the hard-scatter vertex PV$_0$, divided by the fully calibrated jet $p_{\rm{T}}$, i.e.
\vspace{-0.5cm}
\begin{center}
\begin{equation}
\label{eq:RpT}
R_{\rm{p_{T}}}=\frac{\sum_k p_{\rm{T}}^{\rm{trk_k}}(\rm{PV}_0)}{p_{\rm{T}}^{\rm{jet}}}
\end{equation}
\end{center}
Pile-up jets typically have a small value of $R_{\rm{p_{\rm{T}}}}$ corresponding to jets with small charged particle $p_{\rm{T}}$ fraction originating from the hard scatter vertex PV$_0$, as shown in the left plot of Fig.~\ref{fig:rejection-pileup-vs-eff-HS}.

The right plot of Fig.~\ref{fig:rejection-pileup-vs-eff-HS} shows the rejection (defined as the inverse of the mis-tag efficiency) of pile-up jets as a function of the efficiency for selecting hard-scatter jets using the $R_{p_{\rm{T}}}$ discriminant for jets with low and high $p_{\rm{T}}$ in $t\bar{t}$ events with $<\mu>$ = 200 with and without the HGTD, for different time resolution ($\sigma(t)$) values. The extended coverage of ITk makes the track-based pile-up suppression possible in the forward region. A significant improvement in performance of up to a factor of 4 higher pile-up jet rejection at constant efficiency is achieved with the use of timing information.

\section{Bench mark analyses to evaluate the physics performance at HL-LHC}
The effects of an upgraded ATLAS detector are taken into account by applying energy smearing, object efficiencies and fake rates to truth level quantities, and following parameterizations based on detector performance studies with full simulation and HL-LHC conditions. An integrated luminosity of 3000\,fb$^{-1}$ and an average pile-up of $<\mu> = 200$ are assumed.

\subsection{Higgs-boson pair production}
Measurement of the Higgs-boson pair production is a major goal of the HL-LHC.
The small cross section of $\sigma(pp\rightarrow HH)\sim 40$\,fb at $\sqrt{s}=14$\,TeV motivates the use of the dominant gluon fusion production. The channels in which at least one Higgs decaying to $b\bar{b}$ are promising. Thus the high-performance $b$-tagging capability is of critical importance for these analyses. The SM non-resonant $HH$ production process is dominated by gluon-gluon fusion, leading to centrally produced Higgs bosons, hence the extended forward tracking capability of the ITk is not expected to lead to large improvements in sensitivity.

\begin{figure}[h]
\centering
\includegraphics[width=0.46\textwidth]{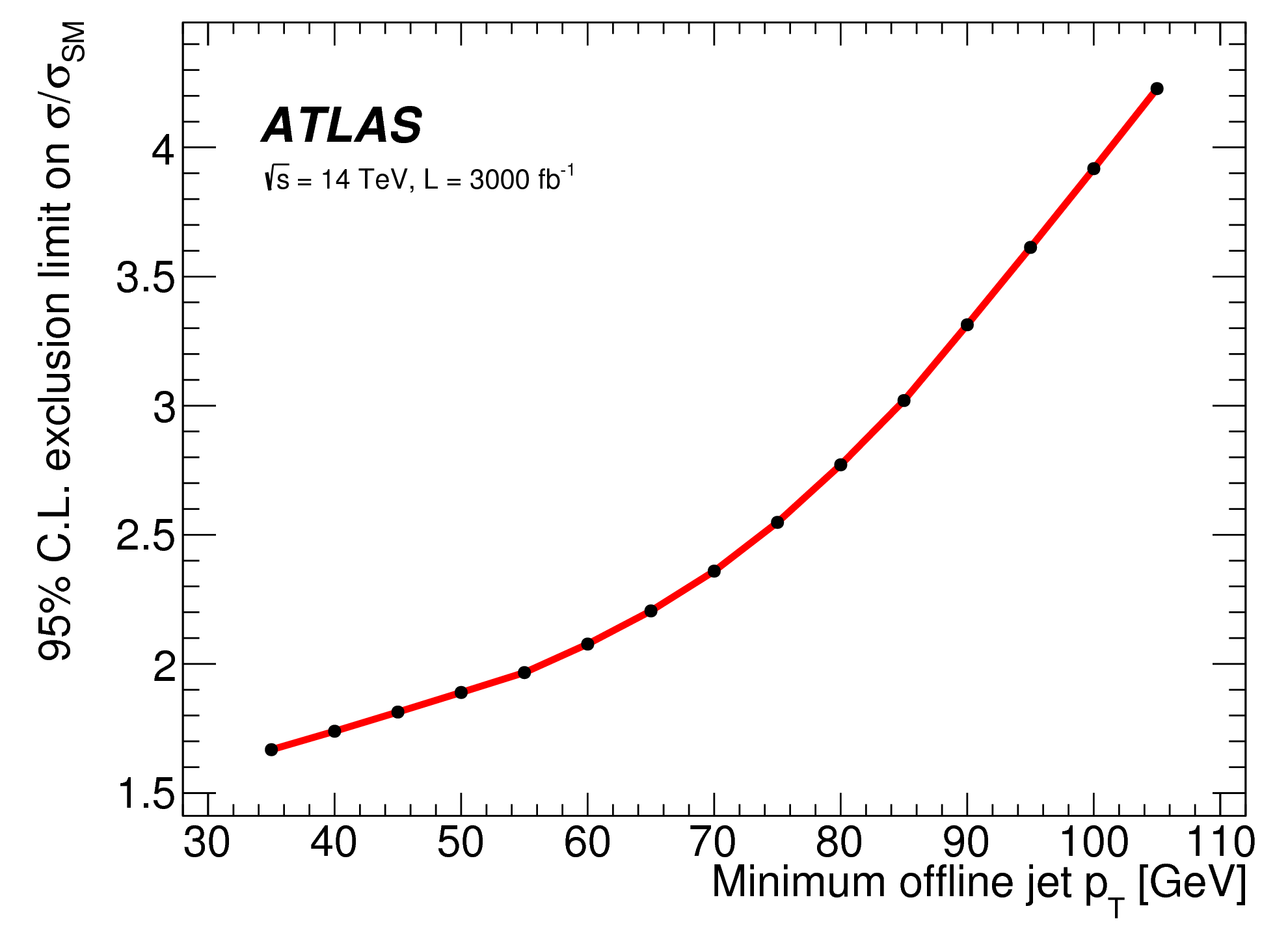}
\hspace{0.5cm}
\includegraphics[width=0.46\textwidth]{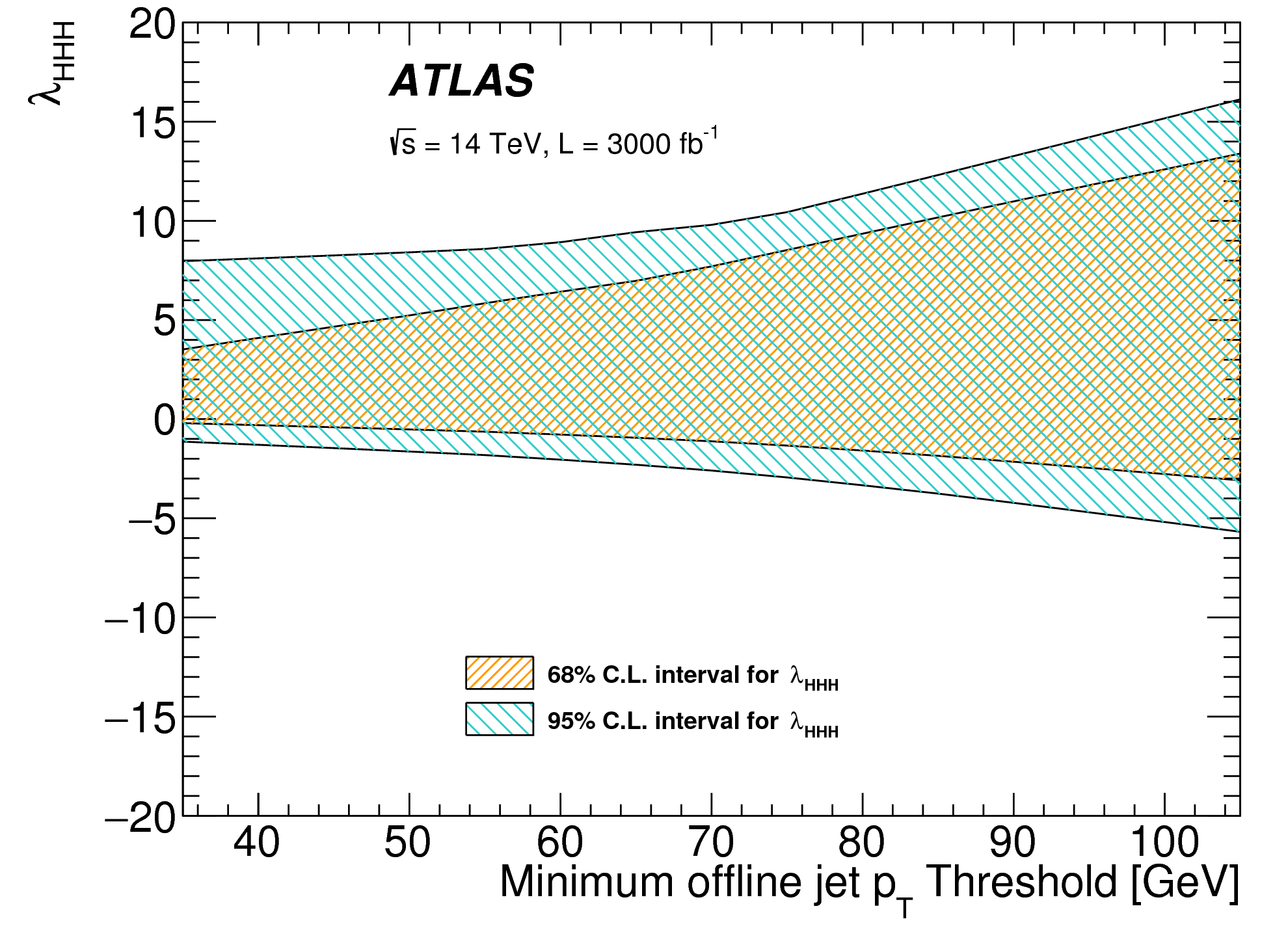}
\caption{Left: Non-resonant $HH\rightarrow 4b$ $\sigma/\sigma_{SM}$ 95\% exclusion sensitivity as a function of the minimum offline jet $p_{\rm{T}}$~\cite{TDAQ-TDR}. Right: Allowed intervals for the $\lambda_{HHH}$ parameter assuming the standard model as function of the minimum offline jet $p_{\rm{T}}$~\cite{TDAQ-TDR}.}
\label{fig:HH-4b}
\end{figure}

The sensitivity of $HH\rightarrow 4b$ channel is estimated as an illustration. The cross section of this channel is predicted to be 13.2\,fb in the SM. This channel has a high sensitivity to $b$-jet trigger threshold. If there is no upgrade to the trigger system, the $p_{\rm{T}}$ threshold for the $b$-jets would be 100\,GeV.
The left plot of Fig.~\ref{fig:HH-4b} shows the 95\% C.L. upper limit on the cross section as a function of the jet $p_{\rm{T}}$ threshold without considering the systematic uncertainty.  Relative to 100\,GeV, the threshold of 65\,GeV which could be reached with the trigger upgrade, will improve the sensitivity by a factor of 2. So the Trigger system upgrade is critical. In the right plot, the upper limit on the Higgs trilinear self-coupling $\lambda_{HHH}$ has been set which is also dependent on the trigger threshold. Based on these results, it is not possible to get a good constraint on SM. More channels are need to be combined to get enough statistics. The combination of the different sub-channels may provide a confirmation of double Higgs production, and perhaps shed some light on the size of $\lambda_{HHH}$.

\subsection{$t\bar{t}$ resonance search}
BSM theories predict new particles with masses at the TeV scale. Specifically a Leptophobic $Z^{'}$ arising from topcolour-assisted technicolour is considered which decays primarily to a pair of top quarks. Thus a $t\bar{t}$ resonance search is a benchmark analysis for evaluating BSM physics prospects at the HL-LHC. The most recent ATLAS search~\cite{Run2:ttbar}, using 36.1\,fb$^{-1}$ of data taken at $\sqrt{s}=13$\,TeV, excludes such $Z^{'}$ bosons for masses less than 3.2\,TeV.

\begin{figure}[h]
\centering
\includegraphics[width=0.46\textwidth]{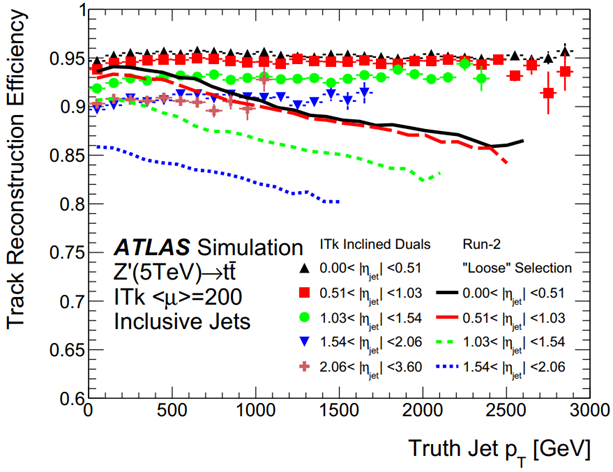}
\hspace{0.5cm}
\includegraphics[width=0.46\textwidth]{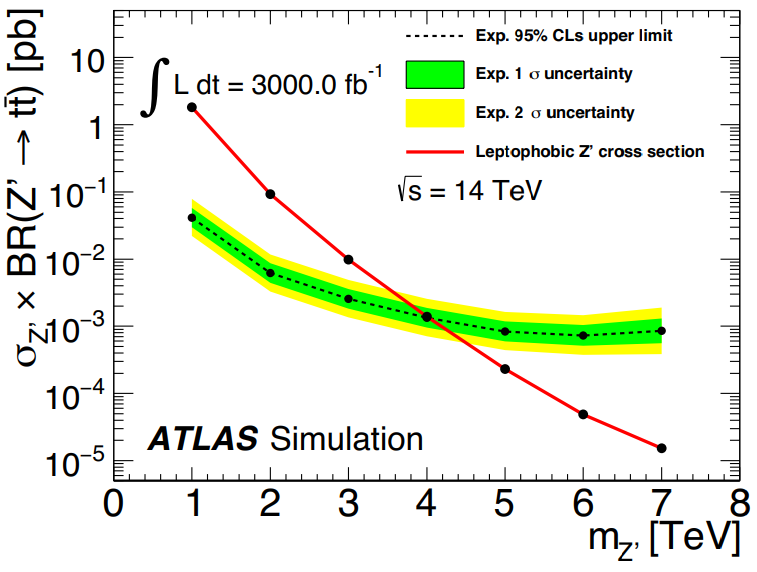}
\caption{Left: Reconstruction efficiency as a function of jet $p_{\rm{T}}$, for tracks in jets from $Z^{'}$(5\,TeV)$\rightarrow t\bar{t}$ events with an average of 200 pile-up events~\cite{pixel-TDR}. Right: The expected upper limit on the cross section $\times$ branching ratio of a leptophobic topcolour-assisted technicolour $Z^{'}$ boson for masses in the range $1-7$\,TeV, assuming an integrated luminosity of 3000\,fb$^{-1}$ at $\sqrt{s}=14$\,TeV~\cite{strip-TDR}.}
\label{fig-Zp}
\end{figure}

At higher masses of the new heavy particles, the produced top quarks will be highly boosted. The top candidates will tend to produce $b$-jets with $p_{\rm{T}} > 600$\,GeV leading to a very dense tracking environment. Thus to be sensitive in this mass region, it is critical to maintain stable track reconstruction efficiency with increasing $p_{\rm{T}}$. Fig.~\ref{fig-Zp} (a) shows the track reconstruction efficiency as a function of jet $p_{\rm{T}}$ for tracks in jets for which $Z^{'}$(5\,TeV)$\rightarrow t\bar{t}$ events with an average of 200 pile-up events have been used. While the current detector shows a significant loss of efficiency for high-$p_{\rm{T}}$ jets is this boosted topology, a similar degradation is not observed for the ITk, thanks to the improved sensor granularity and the enlarged lever arm of the 5 pixel layers.

A topcolour model containing a spin-1 $Z^{'}$ boson is used as the signal assuming a signal width of 1.2\%. Samples at $m_{Z^{'}}=1-7$\,TeV were generated. Leading order cross sections are used for each signal sample, multiplied by a factor of 1.3 to account for NLO effects.  The expected upper limits set on the signal cross section $\times$ branching ratio as a function of the signal mass are shown in the right plot of Fig.~\ref{fig-Zp}. If there is no signals observed, the mass reach of the search is estimated to be 4\,TeV for the leptophobic topcolour-assisted technicolour $Z^{'}$~\cite{ttbar-HL}. The mass reach of $Z^{'}$ is estimated to be 3\,TeV with 300\,fb$^{-1}$. Apparently these estimations are too conservative. The limit on the mass reach is expected to be significantly improved.

\section{Conclusions}
The ATLAS detector upgrade for HL-LHC manages to maintain or improve the performance in very dense environment with pile-up up to 200.
The all-silicon ITk with extended coverage will improve the tracking performance. As a complement of ITk, the new designed timing detector plays important role in the mitigation of pile-up effects. The upgrade to the trigger system is critical to keep lower trigger threshold. With the efforts we have made and the potential improvements in future, the performance of the physics objects reconstruction is expected to be better than the current detector.

\begin{flushleft}

\end{flushleft}


\begin{thebibliography}{11}


\bibitem{ATLAS-detector}
ATLAS Collaboration, 2008 JINST 3 S08003.

\bibitem{pixel-TDR}
ATLAS Collaboration, CERN-LHCC-2017-021, http://cdsweb.cern.ch/record/2285585.

\bibitem{strip-TDR}
ATLAS Collaboration, CERN-LHCC-2017-005, http://cdsweb.cern.ch/record/2257755.

\bibitem{Muon-TDR}
ATLAS Collaboration, CERN-LHCC-2017-017, http://cdsweb.cern.ch/record/2285580.

\bibitem{LAr-TDR}
ATLAS Collaboration, CERN-LHCC-2017-018, http://cdsweb.cern.ch/record/2285582.

\bibitem{Tile-TDR}
ATLAS Collaboration, CERN-LHCC-2017-019, http://cdsweb.cern.ch/record/2285583.

\bibitem{TDAQ-TDR}
ATLAS Collaboration, CERN-LHCC-2017-020, http://cdsweb.cern.ch/record/2285584.

\bibitem{HGTD-TP}
ATLAS Collaboration, CERN-LHCC-2018-023, http://cdsweb.cern.ch/record/2623663.

\bibitem{phaseII-scoping}
ATLAS Collaboration, CERN-LHCC-2015-020, http://cds.cern.ch/record/2055248.

\bibitem{Run2:ttbar}
ATLAS Collaboration, arXiv:1804.10823 [hep-ex].

\bibitem{ttbar-HL}
ATLAS Collaboration, ATL-PHYS-PUB-2017-002, https://cds.cern.ch/record/2243753.

\end{thebibliography}
\end{document}